\documentclass[preprint,showpacs,preprintnumbers,endfloats*,amsmath,amssymb,linenumbers,prb]{revtex4-1}
\usepackage{graphicx}% Include figure files
\usepackage{dcolumn}% Align table columns on decimal point
\usepackage{bm}% bold math
\newcommand{\vecvar}[1]{\mbox{\boldmath$#1$}}
\newcommand{\e}{\mbox{e}}

\begin{document}

\preprint{PRESAT-8501}

\title{First-principles transport calculation method based on real-space finite-difference nonequilibrium Green's function scheme}

\author{Tomoya Ono$^1$, Yoshiyuki Egami$^2$, and Kikuji Hirose$^1$}
\affiliation{$^1$Graduate School of Engineering, Osaka University, Suita, Osaka 565-0871, Japan\\
$^2$Nagasaki University Advanced Computing Center, Nagasaki University, Bunkyo-machi, Nagasaki 852-8521, Japan }

\date{\today}% It is always \today, today,

\begin{abstract}
We demonstrate an efficient nonequilibrium Green's function transport calculation procedure based on the real-space finite-difference method. The direct inversion of matrices for obtaining the self-energy terms of electrodes is computationally demanding in the real-space method because the matrix dimension corresponds to the number of grid points in the unit cell of electrodes, which is much larger than that of sites in the tight-binding approach. The procedure using the ratio matrices of the overbridging boundary-matching technique [Phys. Rev. B {\bf 67}, 195315 (2003)], which is related to the wave functions of a couple of grid planes in the matching regions, greatly reduces the computational effort to calculate self-energy terms without losing mathematical strictness. In addition, the present procedure saves computational time to obtain Green's function of the semi-infinite system required in the Landauer-B\"uttiker formula. Moreover, the compact expression to relate Green's functions and scattering wave functions, which provide a real-space picture of the scattering process, is introduced. An example of the calculated results is given for the transport property of the BN ring connected to (9,0) carbon nanotubes. The wave function matching at the interface reveals that the rotational symmetry of wave functions with respect to the tube axis plays an important role in electron transport. Since the states coming from and going to electrodes show threefold rotational symmetry, the states in the vicinity of the Fermi level, whose wave function exhibits fivefold symmetry, do not contribute to the electron transport through the BN ring.
\end{abstract}

\maketitle
\section{Introduction}
\label{sec:introduction}
Recently, electronic-structure calculations have become an important tool for investigating the physics and chemistry of nanoscale systems with the miniaturization of electronic devices. The electron-transport properties of nanoscale systems have been studied actively because they are of significant importance from both fundamental and practical points of view. Owing to the complexity of the problem, such studies are strongly dependent on the existence of reliable numerical treatments based on first-principles approaches.

A number of first-principles calculation methods for the electron-transport properties of nanoscale systems have been proposed so far. They are roughly categorized into two approaches. One approach uses the nonequilibrium Green's function (NEGF). The relation between conductance and Green's function has been derived within the nonequilibrium Keldysh formalism.\cite{keldysh} This approach has been used extensively in connection with tight-binding models and first-principles methods employing localized basis sets consisting of either atomic orbitals\cite{Brandbyge,Sanvito,Taylor} or Gaussians.\cite{gaussian} In this formalism, by making use of Green's functions with energies of nonreal numbers, the electronic structures of the isolated states in the transition region, which are not easily treated only by the energies of real numbers, can be included into the total charge density of the system. On the other hand, the NEGF method has not been employed with the real-space\cite{chelikowsky,tsdg,book} and plane-wave formalisms so far because the large numbers of grid points or plane waves prevent the direct inversion of $N_x \times N_y \times N_z$-dimensional matrices to obtain the surface Green's functions and self-energy terms of electrodes, where $N_x$, $N_y$, and $N_z$ are the numbers of grid points along the $x$-, $y$-, and $z$-directions in the unit cell of the left or right electrode and electrons flow along the $z$-direction [see, e.g., Fig.~\ref{fig:1}].

The other approach is to compute the scattering wave functions from which transmission coefficients can be obtained. In addition, the scattering wave functions provide a direct real-space picture of the scattering process. This approach has been employed by combining it with techniques where real-space grids and/or plane wave basis sets\cite{lang,tsukada,choi,nkobayashi,tsukamoto,obm,book} are used to describe wave functions and potentials. The easiest way to obtain scattering wave functions is to solve the Lippman-Schwinger equation where the semi-infinite electrode is replaced with a uniformly distributed charge background, i.e., ``jellium.''\cite{lang} The scattering wave function can alternatively be calculated by the wave-function-matching approach proposed by Fujimoto and Hirose, i.e., the overbridging boundary-matching (OBM) method.\cite{obm,book} In the OBM method, several parts of Green's functions of the transition region are computed to set up the wave-function-matching formula and there is no limitation of the usage of jellium, i.e., more realistic atomic and electronic structures of the electrode can be taken into account. Moreover, the effect of electrodes is included in the matching formula as $N_x \times N_y \times \mathcal{N}_f$-dimensional ratio matrices, where $\mathcal{N}_f$ is the order of the finite-difference approximation\cite{chelikowsky} for the kinetic energy operator in the Kohn-Sham equation and is much smaller than $N_z$.

In this paper, we propose the real-space finite-difference (RSFD) NEGF scheme and demonstrate that the surface Green's functions and self-energy terms of the electrodes required in the NEGF method can be obtained from the ratio matrices in the OBM method. The dimension of the matrices for the surface Green's functions and self-energy terms that are directly inverted is reduced from $N_x \times N_y \times N_z$ to $N_x \times N_y \times \mathcal{N}_f$ in the present scheme while keeping the rigorousness of the mathematical formulation. This advantage also saves computational effort to calculate Green's functions in the transition region because the number of elements of the Green's-function matrix required for the transport calculation is proportional to the dimension of the matrices of the self-energy terms. In addition, we prove that scattering wave functions, which help us to interpret the scattering process and are usually calculated by wave function matching methods, can be obtained within the framework of the RSFD NEGF method.

The rest of this paper is organized as follows: in Sec. II, we relate the important quantities of the NEGF method to those of the OBM scheme, and introduce the RSFD NEGF method. In Sec. III, we present an example showing that scattering wave functions help us to interpret the scattering process by computing the transport properties of a BN ring sandwiched between carbon nanotube electrodes (C/BNNT). In Sec. IV, we summarize our procedures. Finally, a mathematical proof is given in appendix.

\section{NEGF method using the computational procedure of the OBM method}
\label{sec:2}
\subsection{Green's function of a whole system including the transition region and two semi-infinite electrodes}
\label{sec:2a}
\begin{figure*}
\begin{center}
\includegraphics{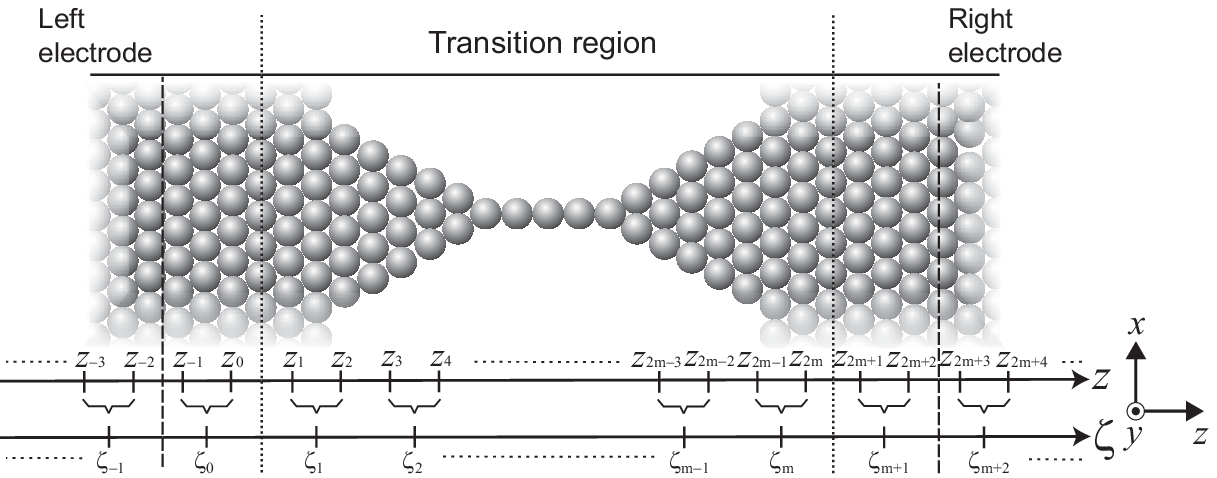}
\caption{Sketch of a system with the transition region intervening between the left and right semi-infinite crystalline electrodes. The dashed lines correspond to the borders of the partitioning of the Hamiltonian matrix in Eq.~(\ref{eqn:sec1-01}) and Fig.~\ref{fig:2}, whereas the dotted lines denote the borders of individual regions used in wave-function matching.\cite{obm} The case for $\mathcal{N}_f=2$ and $N_z=2m$ is illustrated as an example, where $\mathcal{N}_f$ corresponds to the order of the finite-difference approximation for the kinetic energy operator. \label{fig:1}}
\end{center}
\end{figure*}

Let us consider Green's function of a system composed of the transition region sandwiched between two semi-infinite crystalline electrodes, as shown in Fig.~\ref{fig:1}, within the framework of the RSFD scheme. Two-dimensional periodicity in the $x$- and $y$-directions is assumed and a generalized $z$-coordinate $\zeta_l$ instead of $z_l$ is used, because a couple of grid planes are involved in wave-function and Green's-function matching when higher-order finite-difference approximation is employed (see Fig.~\ref{fig:1}). The exchange-correlation effect is treated by the local density approximation\cite{lda} or generalized gradient approximation\cite{gga} of the density functional theory.\cite{dft,kohnsham}

\begin{figure*}[t]
\begin{center}
\includegraphics{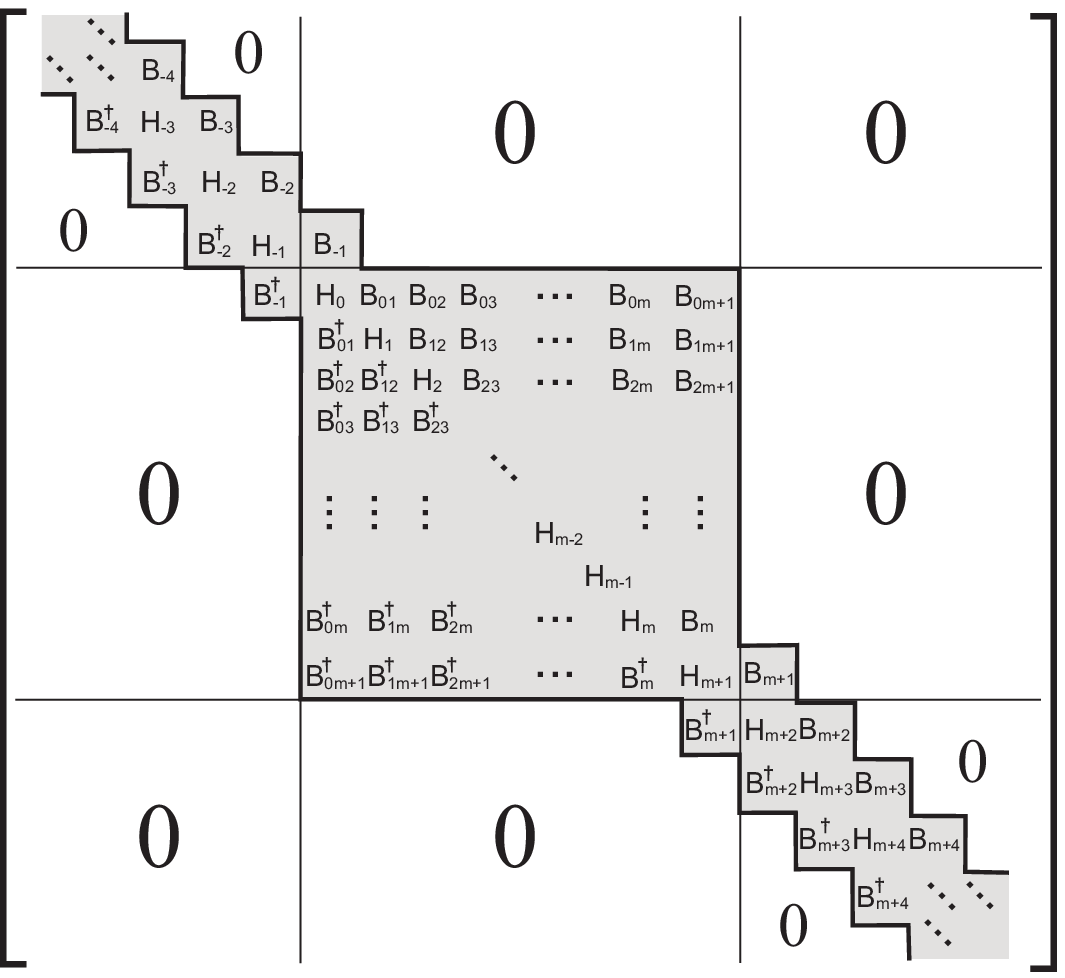}
\caption{Partitioning of the Hamiltonian matrix $\hat{H}$ of Eq.~(\ref{eqn:sec1-01}), associated with the whole system sketched in Fig.~\ref{fig:1}. Block-matrix elements $H_l,B_l$ and $B_{ll'}$ are the abbreviations of $H(\zeta_l,\vecvar{k}_{||})$, $B(\zeta_l)$ and $B(\zeta_l,\zeta_{l'})$, respectively. The partition lines are identical to those in Eq.~(\ref{eqn:sec1-01}). \label{fig:2}}
\end{center}
\end{figure*}

The Green's-function matrix involves the inversion of an infinite matrix corresponding to the Hamiltonian matrix of the whole system $\hat{H}(\vecvar{k}_{||})$, where $\vecvar{k}_{||}$ is the lateral Bloch vector. As shown in Fig.~\ref{fig:2}, we are, however, interested in the finite part of the Green's-function matrix,
\begin{equation}
\hat{H}(\vecvar{k}_{||})=
\left[ 
\begin{array}{c|c|c}
\hat{H}_L(\vecvar{k}_{||}) & \hat{B}_{LT} & 0 \\ 
\hline
 \hat{B}_{LT}^{\dagger} & \hat{H}_T(\vecvar{k}_{||}) & \hat{B}_{TR} \\ 
\hline
0 & \hat{B}_{TR}^{\dagger} & \hat{H}_R(\vecvar{k}_{||})
\end{array}
\right],
\label{eqn:sec1-01}
\end{equation}
where the borders of the partitioning of $\hat{H}(\vecvar{k}_{||})$ are drawn according to the dashed lines in Fig.~\ref{fig:1}; the submatrix $\hat{H}_T(\vecvar{k}_{||})$ contains the matrix elements in the transition region, $\hat{H}_L(\vecvar{k}_{||})$ ($\hat{H}_R(\vecvar{k}_{||})$) corresponds to the semi-infinite left (right)-electrode region, and $\hat{B}_{LT}$ ($\hat{B}_{TR}$), which is nonzero only for some connection area between the transition region and the left (right) electrode, is the interaction between the transition region and the electrodes. $\hat{H}_T(\vecvar{k}_{||})$ in the transition region is treated as a general nonsparse matrix here and in subsequent subsections. On the other hand, $\hat{H}_L(\vecvar{k}_{||})$ ($\hat{H}_R(\vecvar{k}_{||})$) is a block tridiagonal matrix in the RSFD formalism, all of the block-matrix elements of which are $N(=N_x \times N_y \times \mathcal{N}_f)$ dimensional. In addition, $\hat{B}_{LT}$ ($\hat{B}_{TR}$) is a zero matrix except for one $N$-dimensional block-matrix element $B(\zeta_{-1})$ ($B(\zeta_{m+1})$), as illustrated in Fig.~\ref{fig:2}. In practice, $\mathcal{N}_f$ corresponds to the number of $x$-$y$ grid planes involved in the function-matching region $\zeta_l$ since the order of the finite-difference approximation is chosen so as to include the nonlocal region of pseudopotentials in the matching region\cite{ono-al} as well as to obtain sufficiently accurate results with the used grid spacing.

\begin{figure}
\begin{center}
\includegraphics{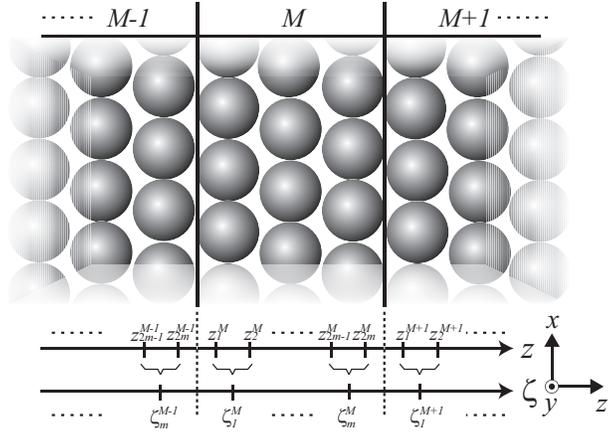}
\caption{Schematic representation of the periodic bulk. $\zeta^M_k$ represents the $z$-coordinate at the $k$th grid plane group in the $M$th unit cell. \label{fig:3}}
\end{center}
\end{figure}

Green's function of the whole system is defined as
\begin{eqnarray}
\hat{G}(Z,\vecvar{k}_{||})&=& \left[ Z-\hat{H}(\vecvar{k}_{||}) \right]^{-1} \nonumber \\
&=& \left[ 
\begin{array}{c|c|c}
\hat{G}_L(Z,\vecvar{k}_{||}) & \hat{G}_{LT}(Z,\vecvar{k}_{||}) & \hat{G}_{LR}(Z,\vecvar{k}_{||}) \\ \hline
\hat{G}_{TL}(Z,\vecvar{k}_{||}) & \hat{G}_T(Z,\vecvar{k}_{||}) & \hat{G}_{TR}(Z,\vecvar{k}_{||}) \\ \hline
\hat{G}_{RL}(Z,\vecvar{k}_{||}) & \hat{G}_{RT}(Z,\vecvar{k}_{||}) & \hat{G}_R(Z,\vecvar{k}_{||})
\end{array}
\right],
\label{eqn:sec1-02}
\end{eqnarray}
where $Z(=E+i\eta)$ is a complex energy variable. From the matrix equation
\begin{equation}
\left[ 
\begin{array}{ccc}
Z-\hat{H}_L(\vecvar{k}_{||}) & -\hat{B}_{LT} & 0 \\ 
-\hat{B}_{LT}^{\dagger} & Z-\hat{H}_T(\vecvar{k}_{||}) & -\hat{B}_{TR} \\ 
0 & -\hat{B}_{TR}^{\dagger} & Z-\hat{H}_R(\vecvar{k}_{||})
\end{array}
\right]
\left[ 
\begin{array}{l}
\hat{G}_{LT}(Z,\vecvar{k}_{||}) \\ 
\hat{G}_{T}(Z,\vecvar{k}_{||}) \\ 
\hat{G}_{RT}(Z,\vecvar{k}_{||}) 
\end{array}
\right]
=
\left[ 
\begin{array}{c}
0 \\ 
I \\ 
0 
\end{array}
\right],
\label{eqn:sec1-03}
\end{equation}
that is,
\begin{eqnarray}
\hat{G}_{LT}(Z,\vecvar{k}_{||})\hat{G}_T(Z,\vecvar{k}_{||})^{-1} = \left[ Z-\hat{H}_L(\vecvar{k}_{||}) \right]^{-1} \hat{B}_{LT} \nonumber \\
-\hat{B}_{LT}^{\dagger}\hat{G}_{LT}(Z,\vecvar{k}_{||}) + \left[ Z-\hat{H}_{T}(\vecvar{k}_{||}) \right] \hat{G}_T(Z,\vecvar{k}_{||})-\hat{B}_{TR}\hat{G}_{RT}(Z,\vecvar{k}_{||})=I \nonumber \\
\hat{G}_{RT}(Z,\vecvar{k}_{||})\hat{G}_T(Z,\vecvar{k}_{||})^{-1} = \left[ Z-\hat{H}_R(\vecvar{k}_{||}) \right]^{-1}\hat{B}_{TR}^{\dagger},
\label{eqn:a04}
\end{eqnarray}
one sees that Green's function of the whole system $\hat{G}_T(Z,\vecvar{k}_{||})$ can be portioned to the transition region as

\begin{equation}
\hat{G}_T(Z,\vecvar{k}_{||}) =\left[ Z-\hat{H}_T(\vecvar{k}_{||})- {\textstyle \hat{\sum}_L}(Z,\vecvar{k}_{||})-{\textstyle \hat{\sum}_R}(Z,\vecvar{k}_{||}) \right]^{-1}.
\label{eqn:sec1-05}
\end{equation}
Note that Eq.~(\ref{eqn:sec1-05}) is equivalent to Dyson's equation in the standard form\cite{dyson} of
\begin{equation}
\hat{G}_T(Z,\vecvar{k}_{||})=\hat{\mathcal{G}}_T(Z,\vecvar{k}_{||})+\hat{\mathcal{G}}_T(Z,\vecvar{k}_{||}) \left[ {\textstyle \hat{\sum}_L}(Z,\vecvar{k}_{||})+{\textstyle \hat{\sum}_R}(Z,\vecvar{k}_{||}) \right] \hat{G}_T(Z,\vecvar{k}_{||}).
\label{eqn:sec1-06}
\end{equation}
Here, ${\textstyle \hat{\sum}_L}(Z,\vecvar{k}_{||})$ and ${\textstyle \hat{\sum}_R}(Z,\vecvar{k}_{||})$ are the self-energy terms of the left and right electrodes defined by
\begin{eqnarray}
{\textstyle \hat{\sum}_L}(Z,\vecvar{k}_{||}) & = & \hat{B}_{LT}^{\dagger} \hat{\mathcal{G}}_L(Z,\vecvar{k}_{||}) \hat{B}_{LT} \nonumber \\
{\textstyle \hat{\sum}_R}(Z,\vecvar{k}_{||}) & = & \hat{B}_{TR} \hat{\mathcal{G}}_R(Z,\vecvar{k}_{||}) \hat{B}_{TR}^{\dagger},
\label{eqn:sec1-07}
\end{eqnarray}
where
\begin{eqnarray}
\hat{\mathcal{G}}_L(Z,\vecvar{k}_{||}) = \left[Z-\hat{H}_L(\vecvar{k}_{||})\right]^{-1} \nonumber \\
\hat{\mathcal{G}}_R(Z,\vecvar{k}_{||}) = \left[Z-\hat{H}_R(\vecvar{k}_{||})\right]^{-1}
\label{eqn:sec1-08}
\end{eqnarray}
are Green's functions of the semi-infinite left and right electrodes with right- and left-side truncations, respectively. In addition, $\hat{\mathcal{G}}_T(Z,\vecvar{k}_{||})$ is the Green's function associated with the isolated transition-region Hamiltonian $\hat{H}_T(\vecvar{k}_{||})$:
\begin{equation}
\hat{\mathcal{G}}_T(Z,\vecvar{k}_{||})=\left[Z-\hat{H}_T(\vecvar{k}_{||})\right]^{-1}.
\label{eqn:sec1-09}
\end{equation}
We used the script capital letter $\hat{\mathcal{G}}$ for describing Green's function of a semi-infinite system with one-side truncation as well as an isolated (two-side truncated) system to prevent confusing ${[Z-\hat{H}_A]}^{-1}$ with $\hat{G}_A$ defined by Eq.~(\ref{eqn:sec1-02}), where $A=L$, $T$ and $R$. From Eqs.~(\ref{eqn:sec1-05}) and (\ref{eqn:sec1-06}), one sees that $\hat{\mathcal{G}}_T(Z,\vecvar{k}_{||})$ is extended to $\hat{G}_T(Z,\vecvar{k_{||}})$ so as to include the effects of semi-infinite electrodes through the self-energy terms ${\textstyle \hat{\sum}}_{\{ L,R \}} (Z,\vecvar{k}_{||})$.

\subsection{Evaluation of self-energy terms}
\label{sec:2b}
This subsection is devoted to the evaluation of the surface Green's functions of the left- and right-electrode regions $\hat{\mathcal{G}}_{\{ L,R \}}(Z,\vecvar{k}_{||})$ and the self-energy terms $\hat{\sum}_{\{L,R\}}(Z,\vecvar{k}_{||})$. Hereafter, we omit the branch of the lateral Bloch vector $\vecvar{k}_{||}$ for simplicity. To set up the Hamiltonian of the electrodes, the Kohn-Sham effective potential is obtained using the unit cell consisting of $N_x \times N_y \times N_z$ grid points under the periodic boundary condition [see Fig.~\ref{fig:3}]. The Green's-function matrices for electrodes are computed using the recursive technique proposed by L\'opez Sancho {\it et al.}\cite{sancho} or Guinea {\it et al.}\cite{guinea} Alternatively, the matrices are evaluated by solving a quadratic eigenvalue problem.\cite{ando} In both schemes, the peculiar characteristic that the $N_x \times N_y \times N_z$-dimensional matrices of the Hamiltonian of the electrodes are the same for each unit cell of the electrodes is employed, which implies that the dimension of the matrices for Green's functions and the self-energy terms becomes $N_x \times N_y \times N_z$. When the number of grid points increases, the computation of these matrices is demanding. In the RSFD NEGF scheme, since $\hat{B}_{LT}$ ($\hat{B}_{TR}$) has only one nonzero $N(=N_x \times N_y \times \mathcal{N}_f)$-dimensional block-matrix element $B(\zeta_{-1})$ ($B(\zeta_{m+1})$) [see Eq.~(\ref{eqn:sec1-01}) and Fig.~\ref{fig:2}], the self-energy terms, which are calculated by Eq.~(\ref{eqn:sec1-07}), are found to take the very simple form of 

\begin{eqnarray}
{\textstyle \hat{\sum}_L}(Z) & = &
\left[ 
\begin{array}{cccc}
{\textstyle \sum _L}(\zeta_0;Z) & 0 & \cdots & 0 \\ 
0 & 0 & \cdots & 0 \\ 
\vdots &  & \vdots &  \\ 
0 & 0 & \cdots & 0 
\end{array}
\right] \nonumber\\
{\textstyle \hat{\sum} _R}(Z) & = &
\left[ 
\begin{array}{cccc}
0 & \cdots & 0 & 0 \\ 
  & \vdots &   & \vdots   \\ 
0 & \cdots & 0 & 0 \\ 
0 & \cdots & 0 & {\textstyle \sum _R}(\zeta_{m+1};Z) 
\end{array}
\right],
\label{eqn:sec2-01}
\end{eqnarray}
where
\begin{eqnarray}
{\textstyle \sum_L}(\zeta_0;Z) & = & B(\zeta_{-1})^{\dagger} \mathcal{G}_L(\zeta_{-1},\zeta_{-1};Z) B(\zeta_{-1}) \nonumber \\
{\textstyle \sum_R}(\zeta_{m+1};Z) & = & B(\zeta_{m+1}) \mathcal{G}_R(\zeta_{m+2},\zeta_{m+2};Z) B(\zeta_{m+1})^{\dagger}
\label{eqn:sec2-02}
\end{eqnarray}
with $\mathcal{G}_{\{L,R\}}(\zeta_k,\zeta_l;Z)$ being the $N$-dimensional $(k,l)$ block-matrix element of $\hat{\mathcal{G}}_{\{L,R\}}(Z)$. In practical calculations, $\mathcal{N}_f$ is much smaller than $N_z$.

To obtain Green's function of the whole system $\hat{G}(Z)$, it is sufficient to calculate $N$-dimensional matrices for the self-energy terms. However, although the Hamiltonians of the $M$th and $M+1$th unit cell are identical, the sliced matrices of the Hamiltonian of the electrodes, $H(\zeta_l^M)$ and $H(\zeta_{l+1}^M)$, are not the same, where $H(\zeta_l^M)$ is the $l$th $N$-dimensional diagonal block-matrix element of the $N_x \times N_y \times N_z$-dimensional Hamiltonian matrices of the electrodes. Thus, we cannot use the two procedures above mentioned.\cite{sancho,guinea,ando} We introduce a computational procedure for obtaining $N$-dimensional matrices of the surface Green's functions and self-energy terms using ratio matrices in the OBM method,\cite{obm} which are computed by solving the generalized eigenvalue problem for the periodic bulk model shown in Fig.~\ref{fig:3}.
\begin{equation}
\Pi_1(Z) \left[
\begin{array}{c}
\Phi_n(\zeta^{M-1}_m,Z) \\
\Phi_n(\zeta^{M+1}_1,Z) 
\end{array}
\right] = \lambda_n(Z)\Pi_2(Z) \left[
\begin{array}{c}
\Phi_n(\zeta^{M-1}_m,Z) \\
\Phi_n(\zeta^{M+1}_1,Z) 
\end{array}
\right],
\label{eqn:sec2-03}
\end{equation}
where
\begin{eqnarray}
\Pi_{1}(Z) 
&=&\left[
\begin{array}{cc}
 \Theta(\zeta^M_m,\zeta^M_1;Z)B(\zeta^M_m)^{\dagger}  & \: \Theta(\zeta^M_m,\zeta^M_m;Z)B(\zeta^M_m) \\
 0  &  I  \\
\end{array}
\right]  \nonumber \\
\Pi_{2}(Z) 
&=&\left[
\begin{array}{cc}
I  &  0  \\
\Theta(\zeta^M_1,\zeta^M_1;Z)B(\zeta^M_m)^{\dagger}  & \: \Theta(\zeta^M_1,\zeta^M_m;Z)B(\zeta^M_m) \\
\end{array}
\right],
\label{eqn:sec2-04}
\end{eqnarray}
and $\Theta(\zeta^M_k,\zeta^M_l;Z)$ is the $N$-dimensional $(k,l)$ block-matrix element of Green's function of the truncated part of the periodic Hamiltonian in the $M$th unit cell. In addition, Eq.~(\ref{eqn:sec2-03}) is the analytically continued equation of Eq.~(19) in Ref.~\onlinecite{obm}. Note that the generalized eigenproblem of Eq.~(\ref{eqn:sec2-03}) suffers numerical error owing to the extremely large and small absolute values of $\lambda(Z)$ in some cases, which prevent us from conducting an accurate computation of the eigenstates $\{ \Phi_n \}$. To avoid this numerical difficulty, we introduce the following ratios of the generalized eigenstates, which are proposed in Ref.~\onlinecite{obm}:
\begin{eqnarray}
&&R^p(\zeta_0;Z)=Q^p(\zeta_{-1};Z)Q^p(\zeta_0;Z)^{-1} \nonumber \\
&&R^q(\zeta_{m+2};Z)=Q^q(\zeta_{m+2};Z)Q^q(\zeta_{m+1};Z)^{-1},
\label{eqn:sec2-05}
\end{eqnarray}
where 
\begin{eqnarray}
Q^p(\zeta_l;Z)&=&\Bigl[\Phi^p_1(\zeta_l;Z),\Phi^p_2(\zeta_l;Z),...,\Phi^p_N(\zeta_l;Z)\Bigr], \nonumber \\
Q^q(\zeta_l;Z)&=&\Bigl[\Phi^q_1(\zeta_l;Z),\Phi^q_2(\zeta_l;Z),...,\Phi^q_N(\zeta_l;Z)\Bigr].
\label{eqn:sec2-06}
\end{eqnarray}
Since the eigenvalues $\lambda_n(Z)$'s for a nonreal $Z$ are divided {\it evenly} into two groups with $|\lambda_n| > 1$ and $|\lambda_n| < 1$,\cite{LeeJoannopoulos} we set up the $N$-dimensional matrix $Q^p(\zeta_l;Z)$ and $Q^q(\zeta_l;Z)$, which gathers the $N$ eigenstates $\{ \Phi^p_n(\zeta_l; Z)\}$ and $\{ \Phi^q_n(\zeta_l; Z)\}$, $n=1,2,...,N$, for a nonreal $Z$ with $|\lambda_n| > 1$ and $|\lambda_n| < 1$, respectively, using the solutions of Eq.~(\ref{eqn:sec2-03}). Then the accuracy of ratio matrices is improved using the following continued-fraction equations in a self-consistent manner [see Eq.~(25) of Ref.~\onlinecite{obm}]:
\begin{eqnarray}
\label{eqn:cfeq}
R^p(\zeta^{M+1}_1;Z) &=&  \Theta(\zeta^M_m,\zeta^M_m;Z) B(\zeta^M_m) \nonumber \\
&&\hspace{-20mm} + \Theta(\zeta^M_m,\zeta^M_1;Z) B(\zeta^M_m)^\dagger \left[ R^p(\zeta^M_1;Z)^{-1} - \Theta(\zeta^M_1,\zeta^M_1;Z) B(\zeta^M_m)^\dagger \right]^{-1} \Theta(\zeta^M_1,\zeta^M_m;Z) B(\zeta^M_m) \nonumber \\
R^q(\zeta^{M+1}_1;Z) &=&  \Theta(\zeta^M_1,\zeta^M_1;Z) B(\zeta^M_m)^\dagger \nonumber \\
&&\hspace{-20mm} + \Theta(\zeta^M_1,\zeta^M_m;Z) B(\zeta^M_m) \left[ R^q(\zeta^{M+1}_1;Z)^{-1} - \Theta(\zeta^M_m,\zeta^M_m;Z) B(\zeta^M_m) \right]^{-1} \Theta(\zeta^M_m,\zeta^M_1;Z) B(\zeta^M_m)^\dagger. \nonumber \\
\label{eqn:sec2-07}
\end{eqnarray}

Now, we prove the following relation that gives a definite description of the surface Green's functions (or self-energy terms) in terms of the ratio matrices of the generalized eigenstates of Eq.~(\ref{eqn:sec2-03}). The surface Green's functions of the left and right electrodes are explicitly expressed as
\begin{eqnarray}
\mathcal{G}_L (\zeta_{-1},\zeta_{-1};Z) & = & R^p(\zeta_{0};Z)B(\zeta_{-1})^{-1} \nonumber \\
\mathcal{G}_R (\zeta_{m+2},\zeta_{m+2};Z) & = & R^q(\zeta_{m+2};Z)B(\zeta_{m+1})^{\dagger -1},
\label{eqn:sec2-08}
\end{eqnarray}
and the self-energy terms Eq.~(\ref{eqn:sec2-02}) are given by
\begin{eqnarray}
{\textstyle \sum_L} (\zeta_{0};Z)&=&B(\zeta_{-1})^{\dagger}R^p(\zeta_{0};Z) \nonumber \\
{\textstyle \sum_R} (\zeta_{m+1};Z) &=&B(\zeta_{m+1}) R^q(\zeta_{m+2};Z).
\label{eqn:sec2-09}
\end{eqnarray}

Hereafter, we concentrate on proving the surface Green's functions and self-energy terms of the left electrode because those of the right electrode can be derived in a similar manner. The proof is derived using the results reported by Lee and Joannopoulos:\cite{LeeJoannopoulos} Green's function $\mathcal{G}_{\{L,R\}}(\zeta_l,\zeta_m;Z\neq real)$ of a crystalline bulk is a decaying function (i.e., $\mathcal{G}_{\{L,R\}} \rightarrow 0$) as $|l| \rightarrow \infty$ with $m$ fixed (or as $|m| \rightarrow \infty$ with $l$ fixed). The eigenstates are $N$-independent functions with respect to $\zeta_l$ with a decreasing or increasing property such that
\begin{eqnarray}
\Phi^p_n(\zeta_{l-sL};Z)&=&(\lambda_n)^{-s} \Phi^p_n(\zeta_{l};Z)
\label{eqn:sec2-10}
\end{eqnarray}
Here, $s$ is an arbitrary positive integer and $L$ is an integer associated with the length of periodicity in the $z$-direction. For example, $L=N_z/\mathcal{N}_f$ when $N_z$ is a multiple of $\mathcal{N}_f$. Furthermore, $\{ \Phi_n(\zeta_l;Z)\}$ satisfies
\begin{eqnarray}
\label{eqn:sec2-11}
-B^{\dagger}_{l-2}\Phi^p_n(\zeta_{l-2};Z)+A_{l-1}\Phi^p_n(\zeta_{l-1};Z)-B_{l-1}\Phi^p_n(\zeta_{l};Z)=0,
\end{eqnarray}
where
\begin{equation}
\label{eqn:sec2-12}
A_l=Z-H(\zeta_l)\ \ \mbox{and}\ \ B_l=B(\zeta_l).
\end{equation}
Note that Eq.~(\ref{eqn:sec2-11}) is the Kohn-Sham equation when a complex number $Z$ is replaced with a real number $E$.

In the left electrode, the surface Green's function $\mathcal{G}_L(\zeta_l,\zeta_{-1};Z)$ is expressed in terms of $Q^p(\zeta_{l};Z)$:
\begin{eqnarray}
\mathcal{G}_L (\zeta_{l},\zeta_{-1};Z) &= &Q^p(\zeta_{l};Z)Q^p(\zeta_{0};Z)^{-1} (B_{-1})^{-1} \hspace{5mm} (l=-1,-2,...,).
\label{eqn:sec2-13}
\end{eqnarray}
In the following, the derivation of Eq.~(\ref{eqn:sec2-13}) is demonstrated. It is straightforward from the definition that $\{ \mathcal{G}_L(\zeta_l,\zeta_{-1};Z)\}$ satisfies
\begin{equation}
\left[ 
\begin{array}{cccccc}
\ddots & \ddots & \ddots & \ddots &   & 0 \\ 
& -B_{-5}^{\dagger}  & A_{-4}  & -B_{-4} &  & \\ 
&  & -B_{-4}^{\dagger}  & A_{-3}  & -B_{-3} & \\ 
&  & & -B_{-3}^{\dagger} & A_{-2}         & -B_{-2} \\ 
0& &  &                & -B_{-2}^{\dagger} & A_{-1}  
\end{array}
\right]
\left[ 
\begin{array}{c}
\vdots \\ 
\mathcal{G}_{-4,-1}  \\ 
\mathcal{G}_{-3,-1}  \\ 
\mathcal{G}_{-2,-1}  \\ 
\mathcal{G}_{-1,-1}
\end{array}
\right]
=
\left[ 
\begin{array}{c}
\vdots \\ 
0  \\ 
0  \\ 
0  \\ 
I
\end{array}
\right],
\label{eqn:sec2-14}
\end{equation}
that is,
\begin{equation}
\begin{array}{ccc}
 \vdots& & \\
-B_{-5}^{\dagger} \mathcal{G}_{-5,-1} + A_{-3} \mathcal{G}_{-4,-1} -B_{-4} \mathcal{G}_{-3,-1} &=& 0 \\
-B_{-4}^{\dagger} \mathcal{G}_{-4,-1} + A_{-3} \mathcal{G}_{-3,-1} -B_{-3} \mathcal{G}_{-2,-1} &=& 0 \\
-B_{-3}^{\dagger} \mathcal{G}_{-3,-1} + A_{-2} \mathcal{G}_{-2,-1} -B_{-2} \mathcal{G}_{-1,-1} &=& 0
\end{array}
\label{eqn:sec2-16}
\end{equation}
and
\begin{equation}
-B_{-2}^{\dagger} \mathcal{G}_{-2,-1} + A_{-1} \mathcal{G}_{-1,-1} = I,
\label{eqn:sec2-15}
\end{equation}
where
\begin{eqnarray}
\mathcal{G}_{l,-1}&=&\mathcal{G}_L (\zeta_l,\zeta_{-1};Z).
\label{eqn:sec2-17}
\end{eqnarray}
From the facts that the $N$-dimensional $\mathcal{G}_L(\zeta_l,\zeta_{-1};Z \ne {\rm real})$ decays deep inside the left electrode ($l \rightarrow -\infty$),\cite{LeeJoannopoulos} we see that $\{ \Phi^p_n(\zeta_l;Z)\}$ exhibits a linear independence of the decaying sequences, where $n=1,2,...,N$, and Eq.~(\ref{eqn:sec2-16}) are the same sets of simultaneous linear equations as Eq.~(\ref{eqn:sec2-11}) for $l \leq -1$. Therefore, $\mathcal{G}_L$ is expanded in terms of $\{ \Phi^p_n \}$ and expressed as
\begin{eqnarray}
\mathcal{G}_L(\zeta_{l},\zeta_{-1};Z) & = & \Biggl[ \sum_{n=1}^N f_{n1} \Phi^p_n(\zeta_{l};Z),\sum_{n=1}^N f_{n2} \Phi^p_n(\zeta_{l};Z),...,\sum_{n=1}^N f_{nN} \Phi^p_n(\zeta_{l};Z) \Biggr] \nonumber \\
&=& Q^p(\zeta_{l};Z) F \hspace{10mm} (l=-1,-2,...),
\label{eqn:sec2-18}
\end{eqnarray}
where $\{ f_{nn'} \}$ is a set of unknown expansion coefficients forming an $N$-dimensional matrix $F$. For simplicity, the dependence of $f_{nn'}$ and $F$ on $Z$ and $\vecvar{k}_{||}$ is ignored. By inserting Eq.~(\ref{eqn:sec2-18}) for $l=-1$ and $-2$ into Eq.~(\ref{eqn:sec2-15}) and subsequently using Eq.~(\ref{eqn:sec2-11}) for $l=0$, we obtain
\begin{equation}
F=Q^p(\zeta_0 ;Z)^{-1} (B_{-1})^{-1}.
\label{eqn:sec2-19}
\end{equation}
By substituting Eq.~(\ref{eqn:sec2-19}) into Eq.~(\ref{eqn:sec2-18}), Eq.~(\ref{eqn:sec2-13}) is obtained.

Next, we carry out the limiting procedure in Eq.~(\ref{eqn:sec2-13}) to obtain the {\it retarded} Green's function. Since the pairing characteristic of $\lambda(E)$'s implies that there are always equal numbers of reflected waves $\Phi^{ref}_n$ and transmitted waves $\Phi^{tra}_n$ for a given energy $E$ and lateral Bloch wave vector $\vecvar{k}_{||}$, we obtain
\begin{eqnarray}
\lim_{\eta\rightarrow 0^{+}}\Phi^p_n(\zeta_l;E+i\eta)&=\Phi^{ref}_n(\zeta_l;E), \nonumber \\
\lim_{\eta\rightarrow 0^{+}}Q^p(\zeta_l;E+i\eta)&=Q^{ref}(\zeta_l;E), \nonumber \\
\lim_{\eta\rightarrow 0^{+}}R^p(\zeta_l;E+i\eta)&=R^{ref}(\zeta_l;E),
\label{eqn:sec2-20}
\end{eqnarray}
where $Q^{ref}(\zeta_l;E)$ is an $N$-dimensional matrix that gathers left-propagating Bloch waves ($|\lambda(E)|=1$ and $\mbox{Re}(\lambda(E))<0$) and rightward increasing evanescent waves ($|\lambda(E)|>1$):
\begin{equation}
Q^{ref}(\zeta_l;E)=\Bigl[\Phi^{ref}_1(\zeta_l;E),\Phi^{ref}_2(\zeta_l;E),...,\Phi^{ref}_N(\zeta_l;E)\Bigr],
\end{equation}
and
\begin{equation}
R^{ref}(\zeta_l;E)=Q^{ref}(\zeta_{l-1};E)Q^{ref}(\zeta_l;E)^{-1}.
\end{equation}
Note that the eigestates with $|\lambda(Z)|=1$ are absent when $\eta \ne 0$,\cite{LeeJoannopoulos} while those with $|\lambda(Z)|=1$ exist in the case of real-number energy.\cite{obm}

Finally, the retarded Green's function is
\begin{eqnarray}
\mathcal{G}_L^r (\zeta_{-1},\zeta_{-1};E) &=& \lim_{\eta \rightarrow 0^{+}} \mathcal{G}_L (\zeta_{-1},\zeta_{-1};E+i\eta) \nonumber \\
& = & Q^{ref}(\zeta_{-1};E)Q^{ref}(\zeta_{0};E)^{-1} (B_{-1})^{-1} \nonumber \\
& = & R^{ref}(\zeta_{0};Z)B(\zeta_{-1})^{-1},
\label{eqn:sec2-21}
\end{eqnarray}
and hence the retarded self-energy term is
\begin{equation}
\textstyle{\hat{\sum}^r_{\{L,R\}}(E)={\displaystyle \lim_{\eta \rightarrow 0^{+}}} \hat{\sum}_{\{L,R\}} (E+i\eta)}.
\label{eqn:sec2-22}
\end{equation} 

Several researchers have investigated the representation of the surface Green's functions by generalized eigenstates.\cite{LeeJoannopoulos,Sanvito,Taylor} Note that the above relations are particularly attractive because they allow us to directly evaluate the retarded surface Green's functions and retarded self-energy terms at a purely real $E$ without using the finite broadening (or smearing) parameter $\eta$. More interestingly, we can reduce the dimension of the matrices of Green's functions and the self-energy terms from $N_x \times N_y \times N_z$ to $N$ while keeping mathematical rigorousness.

\subsection{Evaluation of whole Green's function in the transition region}
\label{sec:2c}
Green's function of the whole system portioned to the transition region, $\hat{G}_T(Z)$, is given by Eq.~(\ref{eqn:sec1-05}). In this subsection, we present the analytic expression of Eq.~(\ref{eqn:sec1-05}), which is equivalent to the exact solution of Dyson's equation Eq.~(\ref{eqn:sec1-06}). In the general case of the Hamiltonian matrix $\hat{H}_T$ being not necessarily sparse, the simplest way to calculate $\hat{G}_T(Z)$ of Eq.~(\ref{eqn:sec1-05}) might be to carry out direct matrix inversion. In practice, however, it is computationally difficult to perform inversion calculations with large matrices. In what follows, we show that there exists an efficient approach to computing the non-Hermit $\hat{G}_T(E)$ on the basis of the OBM scheme.

Let us consider the $l$th column of $\hat{G}_T(Z)$, i.e., $[G_T(\zeta_0,\zeta_l;Z), G_T(\zeta_1,\zeta_l;Z),...,G_T(\zeta_{m+1},\zeta_l;Z)]^t$ ($l=0,1,2,...,m+1$). From Eq.~(\ref{eqn:sec1-05}), one sees that the $l$th column satisfies
\begin{equation}
\left[ Z-\hat{H}_T - {\textstyle \hat{\sum}_L}(Z)-{\textstyle \hat{\sum}_R}(Z) \right]
\left[ 
\begin{array}{c}
G_T(\zeta_0,\zeta_l;Z) \\ 
G_T(\zeta_1,\zeta_l;Z) \\ 
\\
\vdots                 \\
G_T(\zeta_l,\zeta_l;Z) \\
\vdots                 \\
\\
G_T(\zeta_m,\zeta_l;Z) \\ 
G_T(\zeta_{m+1},\zeta_l;Z) \\ 
\end{array}
\right]
=\left[ 
\begin{array}{c}
0 \\ 
0 \\ 
\vdots                 \\ 
0 \\
I  \\
0 \\ 
\vdots                 \\ 
0 \\
0 \\ 
\end{array}
\right]
\begin{array}{l}
 \\ 
 \\ 
 \\ 
 \\ 
\leftarrow \mbox{the} \hspace{2mm} l\mbox{th} \\
 \\ 
 \\ 
 \\ 
 \\ 
\end{array}
\label{eqn:sec3-01}
\end{equation}
by virtue of a simple form of the self-energy matrices, Eq.~(\ref{eqn:sec2-01}). Using Green's function of the truncated part of the Hamiltonian $\hat{\mathcal{G}}_T(Z)$ defined in Eq.~(\ref{eqn:sec1-09}), the whole Green's function in the transition region is given by
\begin{equation}
\left[ 
\begin{array}{c}
G_T(\zeta_0,\zeta_l;Z) \\ 
G_T(\zeta_1,\zeta_l;Z) \\ 
\\
\vdots                 \\
G_T(\zeta_l,\zeta_l;Z) \\
\vdots                 \\
\\
G_T(\zeta_m,\zeta_l;Z) \\ 
G_T(\zeta_{m+1},\zeta_l;Z) \\ 
\end{array}
\right]
=\hat{\mathcal{G}}_T(Z) \left[ 
\begin{array}{c}
\sum_L(\zeta_0;Z)G_T(\zeta_0,\zeta_l;Z) \\ 
0 \\ 
\vdots                 \\ 
0 \\
I  \\
0 \\ 
\vdots                 \\ 
0 \\
\sum_R(\zeta_{m+1};Z)G_T(\zeta_{m+1},\zeta_l;Z) \\ 
\end{array}
\right]
\begin{array}{l}
 \\ 
 \\ 
 \\ 
 \\ 
.\leftarrow \mbox{the} \hspace{2mm} l\mbox{th} \\
 \\ 
 \\ 
 \\ 
 \\ 
\end{array}
\label{eqn:sec3-01a}
\end{equation}
Equation~(\ref{eqn:sec3-01a}) is a matching relation with regard to Green's function $\{G_T(\zeta_k,\zeta_l) \}$, which is an analogous one with regard to the wave function $\{\Psi(z_k)\}$ [see Eq.~(1) of Ref.~\onlinecite{obm}]. The surface Green's-function matching theory has been pioneered by Garc\'\i a-Moliner and Velasco.\cite{Garcia-Moliner and Velasco} From Eq.~(\ref{eqn:sec3-01}), we see that once Green's function of the truncated part of the Hamiltonian $\hat{\mathcal{G}}_T(Z)=[Z-\hat{H}_T]^{-1}$ is known, the elements of the whole Green's function $G_T(\zeta_0,\zeta_l;Z)$ and $G_T(\zeta_{m+1},\zeta_l;Z)$ are calculated using
\begin{eqnarray}
\label{eqn:sec3-02}
\left[
\begin{array}{cc}
\mathcal{G}_T(\zeta_0,\zeta_0;Z){\textstyle \sum_L}(\zeta_0;Z)-I  &\hspace{3mm} \mathcal{G}_T(\zeta_0,\zeta_{m+1};Z){\textstyle \sum_R}(\zeta_{m+1};Z) \\
\mathcal{G}_T(\zeta_{m+1},\zeta_0;Z){\textstyle \sum_L}(\zeta_0;Z)         &\hspace{3mm} \mathcal{G}_T(\zeta_{m+1},\zeta_{m+1};Z){\textstyle \sum_R}(\zeta_{m+1};Z) -I \\
\end{array}
\right] && \nonumber \\
&\hspace{-100mm}\times
\left[
\begin{array}{c}
G_T(\zeta_0,\zeta_l;Z) \\
G_T(\zeta_{m+1},\zeta_l;Z) \\
\end{array}
\right]
=-
\left[
\begin{array}{c}
\mathcal{G}_T(\zeta_0,\zeta_l;Z) \\
\mathcal{G}_T(\zeta_{m+1},\zeta_l;Z) \\
\end{array}
\right].
\end{eqnarray}
Here, $\mathcal{G}_T(\zeta_k,\zeta_l;Z)$ is the $N$-dimensional ($k,l$) block-matrix element of $\hat{\mathcal{G}}_T(Z)$. Equation~(\ref{eqn:sec3-02}), which is a $2N$ simultaneous linear equation with respect to $G_T(\zeta_0,\zeta_l;Z)$ and $G_T(\zeta_{m+1},\zeta_l;Z)$, manifests boundary-value (surface) matching for Green's function of the whole system. To calculate the electronic structure in the transition region, the diagonal elements of the Green's-function matrix, $G_T(\zeta_l,\zeta_l;Z)$, are required. On the other hand, when we are only interested in the transport property, it is sufficient to compute Green's functions on the matching planes, $G^r_T(\zeta_0,\zeta_0;E)$, $G^r_T(\zeta_{m+1},\zeta_0;E)$, $G^r_T(\zeta_0,\zeta_{m+1};E)$, and $G^r_T(\zeta_{m+1},\zeta_{m+1};E)$. In the following, we show the analytic expressions for the solution of Eq.~(\ref{eqn:sec3-02}) in the cases of $l=0$ and $m+1$, which are required in the calculation of the conductance in the NEGF formalism.

For $l=0$ and $l=m+1$,
\begin{eqnarray}
G_T(\zeta_0,\zeta_0;Z) &=& \tilde{\mathcal{G}}_T(\zeta_0,\zeta_0;Z)\left[ I-{\textstyle \sum_L}(\zeta_0;Z)\tilde{\mathcal{G}}_T(\zeta_0,\zeta_0;Z) \right]^{-1}, \nonumber \\
G_T(\zeta_{m+1},\zeta_0;Z) &=& \Bigl[I - \mathcal{G}_T(\zeta_{m+1},\zeta_{m+1};Z){\textstyle \sum_R}(\zeta_{m+1};Z)\Bigr]^{-1} \nonumber \\
&&  \times\ \mathcal{G}_T(\zeta_{m+1},\zeta_{0};Z) \left[ I-{\textstyle \sum_L}(\zeta_0;Z)\tilde{\mathcal{G}}_T(\zeta_0,\zeta_0;Z) \right]^{-1}, \nonumber \\
G_T(\zeta_0,\zeta_{m+1};Z) &=& \Bigl[ I - \mathcal{G}_T(\zeta_{0},\zeta_{0};Z){\textstyle \sum_L}(\zeta_{0};Z)\Bigr]^{-1}\mathcal{G}_T(\zeta_{0},\zeta_{m+1};Z) \nonumber \\ 
& & \times \left[ I-{\textstyle \sum_R}(\zeta_{m+1};Z)\tilde{\mathcal{G}}_T(\zeta_{m+1},\zeta_{m+1};Z) \right]^{-1}, \nonumber \\
G_T(\zeta_{m+1},\zeta_{m+1};Z) &=& \tilde{\mathcal{G}}_T(\zeta_{m+1},\zeta_{m+1};Z)\left[ I-{\textstyle \sum_R}(\zeta_{m+1};Z)\tilde{\mathcal{G}}_T(\zeta_{m+1},\zeta_{m+1};Z) \right]^{-1}, \nonumber \\
\label{eqn:sec3-03}
\end{eqnarray}
where $\tilde{\mathcal{G}}_T$ is a modified $\mathcal{G}_T$ under the influence of the self-energy term ${\textstyle \sum_{\{L,R\}}}$, which is expressed as 
\begin{eqnarray}
\tilde{\mathcal{G}}_T(\zeta_0,\zeta_0;Z) &=& \mathcal{G}_T(\zeta_0,\zeta_0;Z) \nonumber \\
&& \hspace{-20mm} + \mathcal{G}_T(\zeta_0,\zeta_{m+1};Z) {\textstyle \sum_R}(\zeta_{m+1};Z) \nonumber \\
&& \hspace{-20mm} \times\ \Bigl[ I - \mathcal{G}_T(\zeta_{m+1},\zeta_{m+1};Z){\textstyle \sum_R}(\zeta_{m+1};Z)\Bigr]^{-1}\mathcal{G}_T(\zeta_{m+1},\zeta_{0};Z), \nonumber \\
\tilde{\mathcal{G}}_T(\zeta_{m+1},\zeta_{m+1};Z) &=& \mathcal{G}_T(\zeta_{m+1},\zeta_{m+1};Z) \nonumber \\
&& \hspace{-20mm} + \mathcal{G}_T(\zeta_{m+1},\zeta_{0};Z) {\textstyle \sum_L}(\zeta_0;Z) \nonumber \\
&& \hspace{-20mm} \times\ \Bigl[ I - \mathcal{G}_T(\zeta_{0},\zeta_{0};Z){\textstyle \sum_L}(\zeta_{0};Z)\Bigr]^{-1} \mathcal{G}_T(\zeta_{0},\zeta_{m+1};Z). 
\label{eqn:sec3-06}
\end{eqnarray}
It is easy to ensure that the $G_{T}$'s given by Eqs.~(\ref{eqn:sec3-03}) -- (\ref{eqn:sec3-06}) satisfy Eq.~(\ref{eqn:sec3-02}), and therefore, they are the exact analytic solutions of Eq.~(\ref{eqn:sec1-05}) as well as Dyson's equation of Eq.~(\ref{eqn:sec1-06}). Finally, a retarded Green's function is obtainable by carrying out the limiting procedure:
\begin{equation}
G^r_T (\zeta_k,\zeta_l;E) = \lim_{\eta \rightarrow 0^{+}} G_{T} (\zeta_k,\zeta_l;E + i\eta).
\end{equation}

Note that the block matrices of Green's function, $G^r_T(\zeta_0,\zeta_0;E)$, $G^r_T(\zeta_{m+1},\zeta_0;E)$, $G^r_T(\zeta_0,\zeta_{m+1};E)$, and $G^r_T(\zeta_{m+1},\zeta_{m+1};E)$, are not $N_x \times N_y \times N_z$-dimensional, but $N$-dimensional because the matrices of the self-energy terms have already been reduced to $N$-dimensional in the preceding subsection.

\subsection{Description of scattering wave function in terms of whole Green's function}
\label{sec:2d}
We next show that the relationship between the retarded Green's functions and the scattering wave functions is expressed as
\begin{equation}
\Psi_j(\zeta_l;E)=iG_T^r(\zeta_l,\zeta_0;E) \Gamma_L(\zeta_0;E) \Phi_j^{in} (\zeta_0;E) \hspace{20mm}(0 \leq l \leq m+1),
\label{eqn:sec5-05}
\end{equation}
where $\Gamma_L(\zeta_l;E)$ is the coupling matrix, which describes the `coupling strength' of the transition region to the left electrode at $\zeta_0$, and is defined by
\begin{equation}
\Gamma_L(\zeta_0;E)=i\Bigl[ {\textstyle \sum^r_L}(\zeta_0;E)- {\textstyle \sum^a_L}(\zeta_0;E) \Bigr] \hspace{7mm} \left( {\textstyle \sum^a_L}(\zeta_0;E)={\textstyle \sum^r_L}(\zeta_0;E)^{\dagger} \right).
\label{eqn:sec5-04}
\end{equation}

From Eq.~(6) of Ref.~\onlinecite{iobm}, the scattering wave function incoming from deep inside the left electrode is expressed as
\begin{eqnarray}
\left[E-\hat{H}_T-{\textstyle \hat{\sum}^r_L}(E)-{\textstyle \hat{\sum}^r_R}(E)\right]\left[
\begin{array}{c}
\Psi_j(\zeta_{0};E)   \\
\Psi_j(\zeta_{1};E)   \\
\vdots        \\
\Psi_j(\zeta_{m};E)   \\
\Psi_j(\zeta_{m+1};E) \\
\end{array}
\right] && \nonumber \\
&\hspace{-70mm} =
\left[
\begin{array}{c}
B(\zeta_{-1})^{\dagger} \Phi_j^{in}(\zeta_{-1};E)-\sum^r_L(\zeta_0;E)\Phi_j^{in}(\zeta_0;E)   \\ 
    0         \\
\vdots        \\
    0         \\
    0         \\
\end{array}
\right].
\label{eqn:sec5-01}
\end{eqnarray}
The incident wave from the right electrode can be derived in a similar manner. By the definition of the retarded Green's function of the whole system in Eq.~(\ref{eqn:sec1-05}), Eq.~(\ref{eqn:sec5-01}) is rewritten as
\begin{equation}
\left[
\begin{array}{c}
\Psi_j(\zeta_{0};E)   \\
\Psi_j(\zeta_{1};E)   \\
\vdots        \\
\Psi_j(\zeta_{m};E)   \\
\Psi_j(\zeta_{m+1};E) \\
\end{array}
\right] = \hat{G}_T^r(E)
\left[
\begin{array}{c}
B(\zeta_{-1})^{\dagger} \Phi_j^{in}(\zeta_{-1};E)-\sum^r_L(\zeta_0;E)\Phi_j^{in}(\zeta_0;E)   \\ 
    0         \\
\vdots        \\
    0         \\
    0         \\
\end{array}
\right].
\label{eqn:sec5-02}
\end{equation}

Now the ratio matrix $R^{in}$ in the left electrode ($l \le 0$) is introduced along a similar line into the definition of $R^{ref}$:
\begin{equation}
R^{in}(\zeta_l;E)=Q^{in}(\zeta_{l-1};E)Q^{in}(\zeta_l;E)^{-1},
\label{eqn:sec4-02}
\end{equation}
where
\begin{equation}
Q^{in}(\zeta_l;E)=\Bigl[\Phi^{in}_1(\zeta_{l};E),\Phi^{in}_2(\zeta_{l};E),...,\Phi^{in}_N(\zeta_{l};E) \Bigr],
\label{eqn:sec4-03}
\end{equation}
which is assumed to include not only ordinary right-propagating incident Bloch waves but also leftward-decreasing evanescent waves. From the definition of $R^{in}(\zeta_0;E)$, it is straightforward to state that
\begin{equation}
\label{eqn:sec4-04}
\Phi^{in}_j(\zeta_{-1};E)=R^{in}(\zeta_{0};E)\Phi^{in}_j(\zeta_{0};E).
\end{equation}
Furthermore, the relationship between the retarded self-energy term $\sum^r_L(\zeta_{0};E)^{\dagger}$ and $R^{in}(\zeta_0;E)$ is expressed as
\begin{equation}
{\textstyle \sum^r_L}(\zeta_0;E)^{\dagger} = B(\zeta_{-1})^{\dagger} R^{in}(\zeta_0;E)
\label{eqn:sec4-05}
\end{equation}
similarly to that of $\sum^r_L(\zeta_{0};E)$ and $R^{ref}(\zeta_0;E)$ in subsection~\ref{sec:2b}. From Eqs.~(\ref{eqn:sec5-02}), (\ref{eqn:sec4-04}) and (\ref{eqn:sec4-05}), the scattering wave function $\Psi_j(\zeta_l;E)$ for $0 \leq l \leq m+1$ can be written as
\begin{eqnarray}
\Psi_j(\zeta_l;E)&=&G_T^r(\zeta_l,\zeta_0;E) \Bigl[B(\zeta_{-1})^\dagger\Phi_j^{in} (\zeta_{-1};E)-{\textstyle \sum^r_L}(\zeta_0;E) \Phi_j^{in} (\zeta_0;E) \Bigr] \nonumber \\
&=&-G_T^r(\zeta_l,\zeta_0;E) \Bigl[{\textstyle \sum^r_L}(\zeta_0;E)-{\textstyle \sum^r_L}(\zeta_0;E)^\dagger \Bigr] \Phi_j^{in} (\zeta_0;E) \nonumber \\
&=&iG_T^r(\zeta_l,\zeta_0;E) \Gamma_L(\zeta_0;E) \Phi_j^{in} (\zeta_0;E).
\label{eqn:sec5-03}
\end{eqnarray}
The derivation of Eq.~(\ref{eqn:sec5-05}) in a different manner is given in Ref.~\onlinecite{book}, and the expression using $N_x \times N_y \times N_z$-dimensional matrices is also introduced in Ref.~\onlinecite{khomyakov}.

\subsection{Conductance}
\label{sec:2e}
We finally address the problem of electronic transport within the framework of the RSFD Green's-function approach. Here, we consider the case of the incident wave $\Phi^{in}_j$ incoming from deep inside the left electrode. We can prove that the Landauer-B\"uttiker formula\cite{buttiker} $G = {2e^2}/{h} \sum_{i,j} |t_{ij}|^2 {v'_i}/{v_j}$ describing the conductance $G$ has the expression in terms of Green's functions $G^{\{ r,a\} }_T$ and the self-energy matrices $\sum_{ \{ L,R \} }^{ \{ r,a \} }$, where $e$ is the electron charge, $h$ is Planck's constant, $t_{ij}$ is the transmission coefficient for the $j$th incident wave to the $i$th outgoing wave, and $v_j(v'_i)$ is the group velocity  of the state $\Phi^{in}_j(\zeta_0;E)$($\Phi^{tra}_i(\zeta_{m+1};E)$) through the $x$-$y$ plane at $\zeta_0$($\zeta_{m+1}$). In the OBM scheme, group velocity is expressed as
\begin{eqnarray}
v_j &=& L_z \Bigl[{\Phi^{in}_j(\zeta_0;E)}^{\dagger}\Gamma_L(\zeta_0;E)\Phi^{in}_j(\zeta_0;E)\Bigr] \nonumber \\
v'_i &=& L_z \Bigl[{\Phi^{tra}_i(\zeta_{m+1};E)}^{\dagger}\Gamma_R(\zeta_{m+1};E)\Phi^{tra}_i(\zeta_{m+1};E)\Bigr]
\label{eqn:sec4-08}
\end{eqnarray}
where $L_z$ is the length of the unit cell in the $z$-direction. The proof of Eq.~(\ref{eqn:sec4-08}) is given in Appendix~\ref{sec:app1}.

The scattering wave function $\Psi_j(\zeta_l;E)$ corresponding to the $j$th incident wave $\Phi^{in}_j(\zeta_0;E)$ is given by a linear combination of transmitted waves $\Phi^{tra}_i (\zeta_l;E)$ inside the right electrode ($l \ge m+1$) with a transmission coefficient $t_{ij}$, i.e.,
\begin{equation}
\Psi_j(\zeta_l;E) = \sum^{N}_{i=1} t_{ij} \Phi^{tra}_i (\zeta_l;E) 
                    = Q^{tra}(\zeta_l;E) \left[t_{1j},t_{2j},\cdots,t_{Nj}\right]^t,
\label{eqn:sec4-09}
\end{equation}
where $Q^{tra}(\zeta_l;E)$ is an $N$-dimensional matrix that gathers right-propagating Bloch waves ($|\lambda(E)|=1$ and $\mbox{Re}(\lambda(E))>0$) and rightward-decreasing evanescent waves ($|\lambda(E)|<1$) of $\Phi^{tra}_i(\zeta_l;E)$, i.e.,
\begin{equation}
\label{eqn:sec4-10}
Q^{tra}(\zeta_l;E)=\Bigl[\Phi^{tra}_1(\zeta_l;E),\Phi^{tra}_2(\zeta_l;E),\cdots,\Phi^{tra}_N(\zeta_l;E)\Bigr].
\end{equation}
From Eqs.~(\ref{eqn:sec5-05}) and (\ref{eqn:sec4-09}) for $l=m+1$, we have
\begin{equation}
\left[t_{1j},t_{2j},\cdots,t_{Nj}\right]^t = i Q^{tra}(\zeta_{m+1};E)^{-1}G^{r}_T (\zeta_{m+1},\zeta_0;E) \Gamma_L(\zeta_0;E)\Phi^{in}_{L,j}(\zeta_0;E)\label{egami:eqn9-112}
\end{equation}
and then obtain
\begin{equation}
T = i Q^{tra}(\zeta_{m+1};E)^{-1}G^{r}_T (\zeta_{m+1},\zeta_0;E) \Gamma_L(\zeta_0;E)Q^{in}(\zeta_0;E).
\label{eqn:sec4-11}
\end{equation}
Here, $T$ is the transmission-coefficient matrix
\begin{equation}
T=\left[
\begin{array}{cccc}
t_{11} & t_{12} & \cdots & t_{1N} \\
t_{21} & t_{22} & \cdots & t_{2N} \\
       & \cdots &        &        \\
t_{N1} & t_{N2} & \cdots & t_{NN} \\
\end{array}
\right]
\label{eqn:sec4-12}
\end{equation}
and $Q^{in}(\zeta_0;E)$ is the matrix defined by Eq.~(\ref{eqn:sec4-03}). Note that the expression
\begin{equation}
\sum_{i,j}|t_{ij}|^2 \frac{v'_i}{v_j} = \mbox{Tr} \Bigl[ \mathcal{V}^{-1} T^{\dagger} \mathcal{V}' T \Bigr]
\label{eqn:sec4-13}
\end{equation}
holds, with $\mathcal{V}^{(')}$ being a diagonal matrix whose elements are $v_i^{(')}\delta_{ij}$. By substituting Eq.~(\ref{eqn:sec4-11}) into the right-hand side of Eq.~(\ref{eqn:sec4-13}), we find
\begin{eqnarray}
\sum_{i,j}|t_{ij}|^2 \frac{v'_i}{v_j} &=& \mbox{Tr} \Bigl[ \Gamma_L(\zeta_0;E)Q^{in}(\zeta_0;E)\mathcal{V}^{-1}Q^{in}(\zeta_0;E)^{\dagger}\Gamma_L(\zeta_0;E)G^a_T(\zeta_0,\zeta_{m+1};E) \nonumber \\
& & \times\ Q^{tra}(\zeta_{m+1};E)^{\dagger -1} \mathcal{V}'  Q^{tra}(\zeta_{m+1};E)^{-1} G^r_T(\zeta_{m+1},\zeta_0;E) \Bigr].
\label{eqn:sec4-14}
\end{eqnarray}
Here, `Tr' stands for the trace, i.e., the sum of the diagonal matrix elements, and the cyclic property of the trace is used. From Eqs.~(\ref{eqn:sec4-03}), (\ref{eqn:sec4-08}), and (\ref{eqn:sec4-10}), the relations
\begin{eqnarray}
\mathcal{V} &=& iL_z Q^{in}(\zeta_0;E)^{\dagger} \Gamma_L(\zeta_0;E) Q^{in}(\zeta_0;E) \nonumber \\
\mathcal{V}' &=& iL_z Q^{tra}(\zeta_{m+1};E)^{\dagger} \Gamma_R(\zeta_{m+1};E) Q^{tra}(\zeta_{m+1};E)
\label{eqn:sec4-15}
\end{eqnarray}
are derived, and then
\begin{eqnarray}
G &=& \frac{2e^2}{h} \sum_{i,j} |t_{ij}|^2 \frac{v'_i}{v_j} \nonumber \\
  &=& \frac{2e^2}{h} \mbox{Tr} \Bigl[ \Gamma_L(\zeta_0;E)G^a_T(\zeta_0,\zeta_{m+1};E)\Gamma_R(\zeta_{m+1};E)G^r_T(\zeta_{m+1},\zeta_0;E) \Bigr] \nonumber \\
  &=& \frac{2e^2}{h} \mbox{Tr} \Bigl[ \Gamma_L(\zeta_0;E)G^r_T(\zeta_0,\zeta_{m+1};E)\Gamma_R(\zeta_{m+1};E)G^a_T(\zeta_{m+1},\zeta_0;E) \Bigr]
\label{eqn:sec4-16}
\end{eqnarray}
are established. Here, the advanced Green's function is
\begin{equation}
\label{eqn:sec4-17}
G^a_T(\zeta_k,\zeta_l;E)=G^r_T(\zeta_l,\zeta_k;E)^{\dagger}.
\end{equation}
Equation~(\ref{eqn:sec4-16}) is a well-known formula\cite{fisherlee} in the NEGF formalism pioneered by Keldysh.\cite{keldysh} The equality of the last line in Eq.~(\ref{eqn:sec4-16}) is also verified with a similar consideration of the case of incident waves incoming from the right electrode. One of the advantages of the Green's-function approach is that the conductance is calculated without the knowledge of well-defined asymptotic wave functions in the transition region.

\section{Transport property of BN ring connected to CNT electrodes}
\begin{figure}
\begin{center}
\includegraphics{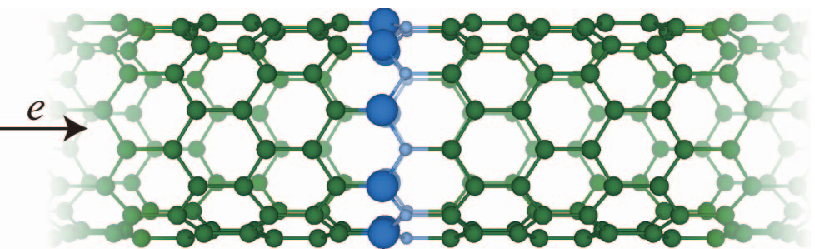}
\caption{(color online) Computational model where one BN zigzag ring is sandwiched between (9,0) CNT electrodes. Large dark, small dark, and small light balls are N, C, and B atoms, respectively. The rectangle enclosed by broken lines represents the supercell used to evaluate the optimized atomic configuration and Kohn-Sham effective potential. \label{fig:4}}
\end{center}
\end{figure}

To demonstrate the applicability of the RSFD NEGF method and the importance of the interpretation using scattering wave functions, the transport property of the C/BNNT where one carbon ring of (9,0) CNT is replaced with a BN ring is examined. Figure~\ref{fig:4} shows the computational model, in which the C/BNNT is sandwiched between the CNT electrodes. A valence electron-ion interaction is described using norm-conserving pseudopotentials\cite{norm} generated by the scheme proposed by Troullier and Martins.\cite{tm} Exchange and correlation effects are treated within the local density approximation\cite{lda} of the density functional theory. To determine the Kohn-Sham effective potential, we use a conventional supercell under a periodic boundary condition in all directions with a real-space grid spacing of $\sim$ 0.24 \AA; the dimensions of the supercell are $L_x=13.34$ \AA, $L_y=13.34$ \AA, and $L_z=4.32$ \AA, where $L_x$ and $L_y$ are the lateral lengths of the supercell in the $x$- and $y$-directions perpendicular to the nanotube axis, respectively, and $L_z$ is the length in the $z$-direction. Then we compute the scattering wave functions obtained non-self-consistently. It has been reported that this procedure is just as accurate in the linear response regime but significantly more efficient than performing computations self-consistently on a scattering-wave basis.\cite{kong2} The conductance is calculated using Eq.~ (\ref{eqn:sec4-16}).

\begin{figure}
\includegraphics{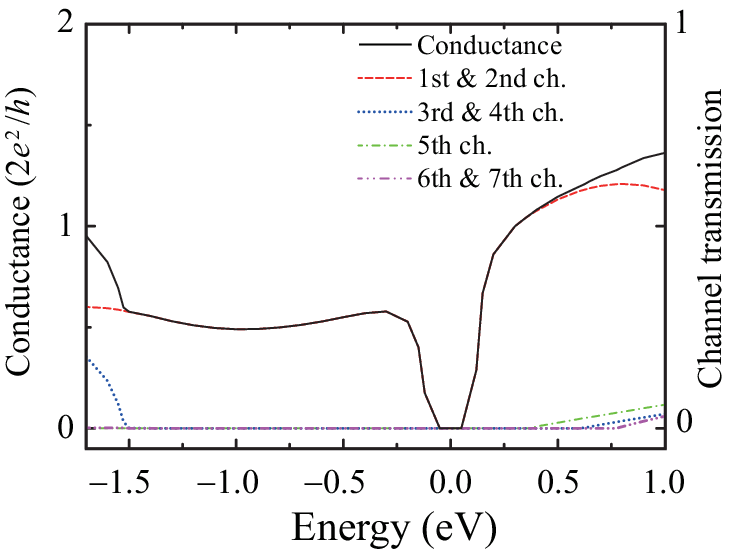}
\caption{Conductance spectra of C/BNNT as functions of energy of incident electrons. The solid curve represents conductance. Dashed, dotted, dash-dotted, and dashed double-dotted curves show channel transmissions. \label{fig:5}}
\end{figure}

\begin{figure}
\includegraphics{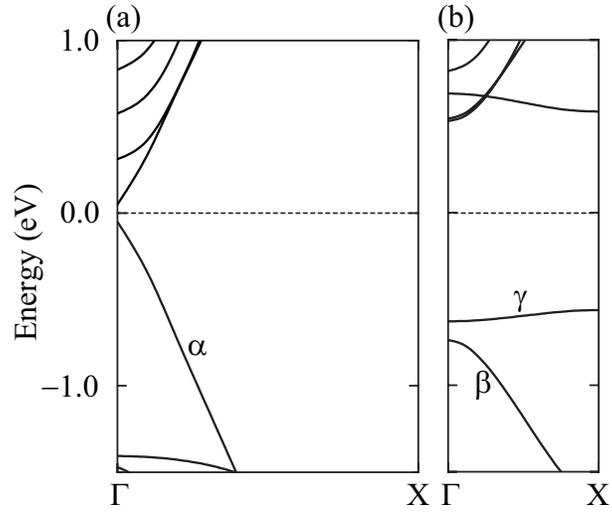}
\caption{Electronic band structures of (a) CNT and (b) C/BNNT. The supercells include two and four (9,0) rings for CNT and C/BNNT, respectively. The zero of energy is the Fermi energy.}
\label{fig:6}
\end{figure}

\begin{figure}
\includegraphics{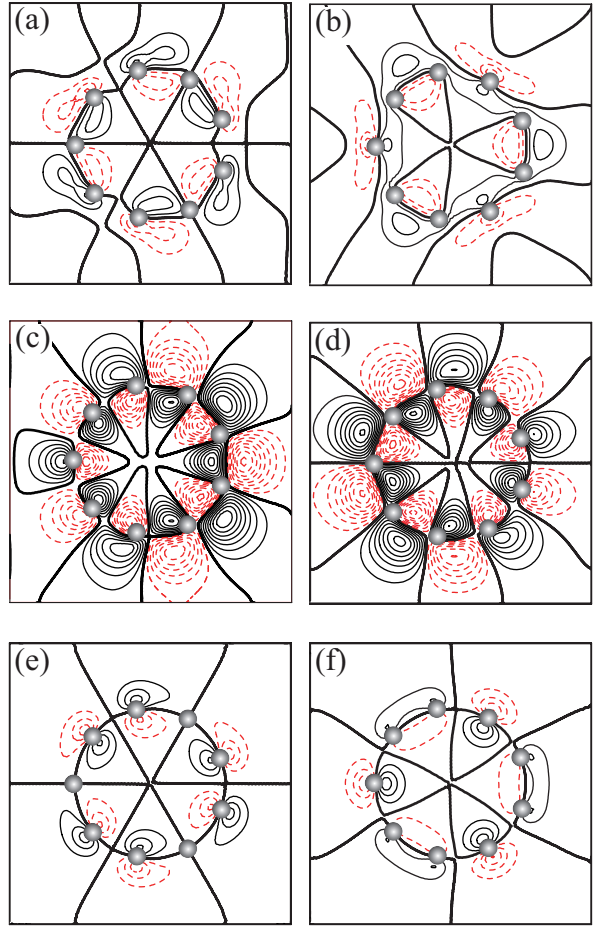}
\caption{Contour plots of wave functions. (a) and (b) Doubly degenerate bands indicated by $\beta$ of C/BNNT in Fig.~\ref{fig:6}. (c) and (d) Doubly degenerate bands indicated by $\gamma$ of C/BNNT. (e) and (f) Doubly degenerate bands indicated by $\alpha$ of CNT. Negative values are indicated by dashed curves; the thick solid curve represents zero. The contour plot is separated by 7.41 $\times 10^{-4}$ electron/\AA$^3$. The spheres represent the position of carbon atoms.}
\label{fig:7}
\end{figure}

\begin{figure}
\includegraphics{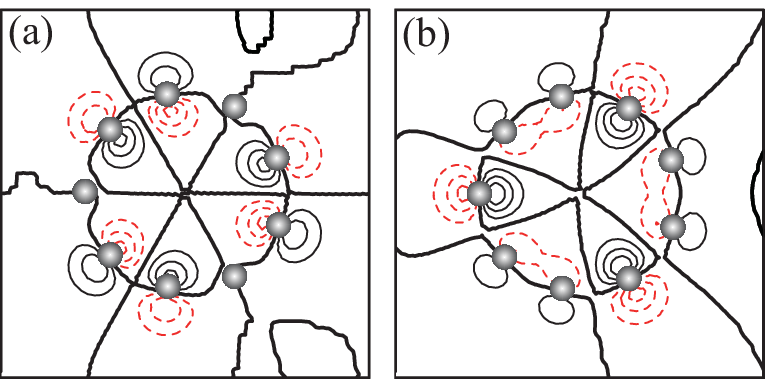}
\caption{Contour plots of doubly degenerate scattering wave functions at energy of $E_F-0.5$ eV. Negative values are indicated by dashed curves; the thick solid curve represents zero. The contour plot is separated by 3.27 $\times 10^{-4}$ electron/\AA$^3$/eV. The spheres represent the position of carbon atoms.}
\label{fig:8}
\end{figure}

Figure~\ref{fig:5} shows the conductance spectrum of the C/BNNT. To investigate states that actually contribute to electron transport, the eigenchannels are computed by diagonalizing the Hermit matrix $T^{\dagger}T$, where $T$ is the transmission-coefficient matrix defined in Eq.~(\ref{eqn:sec4-12}).\cite{nkobayashi} For reference, we plot in Fig.~\ref{fig:6} the band structures of the CNT and C/BNNT in the periodic supercell used to obtain the Kohn-Sham effective potential. It is reported in the other DFT calculations that the band gap of the (9,0) CNT opens even though the (9,0) CNT is expected to be metallic by the zone folding of the tight-binding approximation;\cite{zolyomi} the zero-conductance region around the Fermi level corresponds to the fundamental band gap of the (9,0) CNT. In addition, the bands of the C/BNNT indicated by $\beta$ and $\gamma$ doubly degenerate as well as the bands of the CNT indicated by $\alpha$ at approximately the Fermi level. One can see in Fig.~\ref{fig:5} that two channels actually contribute to the transport in the vicinity of the Fermi level and significant peaks due to the resonant tunneling through the dispersionless bands indicated by $\gamma$ are not observed in the conductance spectrum. Since a real-space picture of the wave function helps us to understand the transport phenomenon, the spatial behaviors of the C/BNNT and CNT states, which are relevant to the electron transport, are shown in Fig.~\ref{fig:7}. We plot the wave functions at $\Gamma$ point because the symmetry in the $x$-$y$ plane is insensitive with respect to the variation in $k_z$ in a one-dimensional Brillouin zone. For comparison, the behaviors of the wave functions indicated by $\alpha$, $\beta$, and $\gamma$ in Fig.~\ref{fig:6} are also plotted in Fig.~\ref{fig:8}. The wave function of the energetically dispersive bands of the C/BNNT (indicated by $\beta$) shows threefold rotational symmetry with respect to tube axis, whereas that of the dispersionless bands (indicated by $\gamma$) shows fivefold symmetry. The energetically dispersive bands can contribute to electron transport because the spatial symmetry of the wave functions of the bands corresponds to that of the wave functions coming from the left CNT electrode and outgoing to the right CNT electrode indicated by $\alpha$. In contrast, the symmetry of the dispersionless bands does not agree with those of the coming and outgoing waves, and thus the wave functions from the CNT electrodes are hardly connected to the wave function of the dispersionless bands, resulting in a small contribution to the electron transport. These results imply that the interpretation using the scattering wave function, which does not explicitly appear in the NEGF method, is important for the investigation of the transport properties of nanoscale systems.

\section{Summary}
We have proposed an efficient procedure for RSFD NEGF calculations for obtaining the transport property of nanoscale systems. Since the number of grid points in real-space methods is larger than that of basis in the tight-binding approach, the computational cost to obtain the surface Green's functions and self-energy terms has been the bottleneck. Using the ratio matrices in the OBM method, the present procedure greatly reduces the computational cost to obtain the surface Green's functions and self-energy terms of the electrodes as well as the Green's functions of the whole system without loss of mathematical rigorousness. In addition, we proved that scattering wave functions, which provide us a real-space picture of the scattering process, can be obtainable by the present scheme.

The transport property of the BN ring connected to the carbon nanotube electrode is investigated using the present method. By examining the rotational symmetry of wave functions at the matching plane, we found that states that are energetically dispersionless in a one-dimensional Brillouin zone hardly contribute to the electron transport because of the difference in the rotational symmetry of wave functions with respect to the tube axis. This result indicates that the real-space picture of the scattering wave function, which is not necessary to be taken into account to calculate the transport property in the NEGF formalism, helps us to interpret transport phenomena in the transition region.

Since the OBM method is developed for the RSFD scheme, the present technique allows us to efficiently obtain quantities for the NEGF method in a real-space representation. In particular, the RSFD scheme of first-principles calculations is a method that has the advantage to scale with massively parallel architectures and has this potential without compromise on precision. Moreover, this scheme is free of problems concerning the completeness of the basis set such as in the methods using localized basis sets of either atomic orbitals or Gaussians. Therefore, the present procedure opens the possibility for executing large-scale RSFD NEGF calculations using massively parallel computers with a high degree of accuracy.

\section*{Acknowledgements}
This research was partially supported by Strategic Japanese-German Cooperative Program from Japan Science and Technology Agency and Deutsche Forschungsgemeinschaft, by the Computational Materials Science Initiative (CMSI), and by a Grant-in-Aid for Scientific Research on Innovative Areas (Grant No. 22104007) from the Ministry of Education, Culture, Sports, Science and Technology, Japan. The numerical calculation was carried out using the computer facilities of the Institute for Solid State Physics at the University of Tokyo and Center for Computational Sciences at University of Tsukuba.

\newpage
\appendix
\section{Group velocity}
\label{sec:app1}
For completeness, we introduce the expression for the group velocity in this appendix. Group velocity is written as
\begin{equation}
\label{eqn:app101}
v_g=\frac{\partial E}{\partial k(E)}\vecvar{u}^\dagger(E)\vecvar{u}(E),
\end{equation}
where $k(E)$ is a Bloch wave vector with $E$ being the Kohn-Sham energy and $\vecvar{u}(E)$ is an $N_x \times N_y \times N_z$-dimensional columnar vector consisting of $\Phi_j(\zeta^M_l,E)$ of the $M$th unit cell (see Fig.~\ref{fig:2}):
\begin{equation}
\label{eqn:app102}
\vecvar{u}(E)=[\Phi_j(\zeta^M_1;E),\Phi_j(\zeta^M_2;E),\cdots,\Phi_j(\zeta^M_N;E)]^t.
\end{equation}
The propagating wave obeys the Kohn-Sham equation
\begin{equation}
\label{eqn:app103}
\hat{H}^M_{per}(k(E),\vecvar{k}_{||})\vecvar{u}(E)=E \vecvar{u}(E),
\end{equation}
where $\hat{H}^M_{per}(k(E)$ is the $N_x \times N_y \times N_z$-dimensional {\it periodic} Hamiltonian of the $M$th unit cell,
\begin{eqnarray}
\label{eqn:app104}
\hat{H}^M_{per}(k(E),\vecvar{k}_{||}) \hspace{90mm} \nonumber \\
= \left[
\begin{array}{cccccc}
H(\zeta^M_{1};\vecvar{k}_{||})  & B(\zeta^M_1)    &     0        &    \cdots     & 0   &     \e^{-i{k(E)}{L_z}} B(\zeta_m^M)^{\dagger}    \\
B(\zeta^M_1)^{\dagger}         & H(\zeta^M_2;\vecvar{k}_{||})   & B(\zeta^M_2)         &        &      &    0      \\
   0         & \ddots     &  \ddots     &  \ddots              &        & \vdots   \\
   \vdots    & & \ddots     &  \ddots     &  \ddots                &  0  \\
   0         &            & B(\zeta^M_{m-2})^{\dagger}     &    & H(\zeta^M_{m-1};\vecvar{k}_{||}) & B(\zeta^M_{m-1})      \\
\e^{i{k(E)}{L_z}} B(\zeta_m^M)  &0   & \cdots      & 0&  B(\zeta^M_{m-1})^{\dagger}      & H(\zeta^M_m;\vecvar{k}_{||})
\end{array}
\right]. \nonumber \\
\end{eqnarray}

Differentiating the eigenvalue equation of Eq.~(\ref{eqn:app103}) with respect to $k(E)$ and multiplying $\vecvar{u}(E)^{\dagger}$ from the left-hand side of the resultant equation yields
\begin{equation}
\label{eqn:app105}
\vecvar{u}(E)^{\dagger}\frac{d\hat{H}^M_{per}}{dk(E)}\vecvar{u}(E)=\frac{dE}{dk(E)}\vecvar{u}(E)^{\dagger}\vecvar{u}(E).
\end{equation}
From Eq.~(\ref{eqn:app104}), one finds that
\begin{equation}
\label{eqn:app106}
\frac{d\hat{H}_{per}}{dk(E)}=\left[
\begin{array}{ccccccc}
0&&0&\cdots&0&&-i L_z  \e^{-i{k(E)}{L_z}} {B(\zeta^M_m)}^{\dagger} \\
0&&0&\cdots&0&&0 \\
\vdots&&\vdots& &\vdots&&\vdots \\
i L_z  \e^{i{k(E)}{L_z}} B(\zeta^M_m)&&0&\cdots&0&&0
\end{array}
\right]
\end{equation}
and 
\begin{eqnarray}
\label{eqn:app107}
{\vecvar{u}(E)}^{\dagger}\frac{d\hat{H}_{per}}{dk(E)}\vecvar{u}(E)&=&iL_z\e^{i{k(E)}{L_z}} {\Phi_j(\zeta^M_m;E)}^{\dagger}B(\zeta^M_m)\Phi_j(\zeta^M_1;E) \nonumber \\
&&-iL_z\e^{-i{k(E)}{L_z}}{\Phi_j(\zeta^M_1;E)}^{\dagger}{B(\zeta^M_m)}^{\dagger}\Phi_j(\zeta^M_m;E) \nonumber \\
&=&iL_z{\Phi_j(\zeta^M_m;E)}^{\dagger}B(\zeta^M_m)\Phi_j(\zeta^{M+1}_1;E) \nonumber \\
&&-iL_z{\Phi_j(\zeta^{M+1}_1;E)}^{\dagger}{B(\zeta^M_m)}^{\dagger}\Phi_j(\zeta^M_m;E) \nonumber \\
&=&iL_z{\Phi_j(\zeta^{M-1}_m;E)}^{\dagger}B(\zeta^M_m)\Phi_j(\zeta^M_1;E) \nonumber \\
&&-iL_z{\Phi_j(\zeta^M_1;E)}^{\dagger}{B(\zeta^M_m)}^{\dagger}\Phi_j(\zeta^{M-1}_m;E).
\end{eqnarray}
Here, the second step follows from the Bloch condition
\begin{equation}
\label{eqn:app108}
\Phi_j(\zeta^{M+1}_1;E)=\e^{i{k(E)}{L_z}} \Phi_j(\zeta^M_1;E),
\end{equation}
and the flux conservation is used in the last step. From Eqs.~(\ref{eqn:app101}), (\ref{eqn:app105}) and (\ref{eqn:app107}), group velocity is given by
\begin{equation}
\label{eqn:app109}
v_g =i L_z \left[ \Phi_j(\zeta^{M-1}_m;E)^\dagger B(\zeta^M_m) \Phi_j(\zeta^{M}_1;E) - \Phi_j(\zeta^M_1;E)^\dagger {B(\zeta^M_m)}^{\dagger} \Phi_j(\zeta^{M-1}_m;E) \right],
\end{equation}
where the normalization of the Bloch states is defined as $\sum_l \Phi_j(\zeta^M_l;E)^\dagger \Phi_j(\zeta^M_l;E)=1$.

Furthermore, making use of Eqs.~(\ref{eqn:sec5-04}), (\ref{eqn:sec4-04}), and (\ref{eqn:sec4-05}), we have a simpler expression for group velocity in terms of the coupling matrix:
\begin{equation}
\label{eqn:app110}
v_g= L_z \left[ \Phi_j(\zeta^{M}_1;E)^\dagger \Gamma(\zeta^{M}_1;E) \Phi_j(\zeta^{M}_1;E) \right].
\end{equation}

\end{document}